\documentclass[%
 reprint,
 amsmath,amssymb,
 aps,
]{revtex4-2}

\usepackage{graphicx}
\usepackage{dcolumn}
\usepackage{bm}

\begin{document}

\preprint{APS/123-QED}

\title{\textit{Ab initio} investigation of a spin-polarized two-dimensional electron gas at the BaTiO$_{3}$/LaMnO$_{3}$ interface}

\author{I.I.~Piyanzina}
\author{R.F.~Mamin}%
\affiliation{%
 Zavoisky Physical-Technical Institute, FIC KazanSC of
	RAS, 420029 Kazan, Russia\\
}%




\date{\today}

\begin{abstract}
By means of \textit{ab initio} calculations the possibilities of switchable spin-polarized two-dimensional electron gas (2DEG) at the interface of antiferromagnetic and ferroelectric perovskites, i.e., LaMnO$_{3}$/BaTiO$_{3}$ superlattice, were investigated. We demonstrate that at the heterostructure with TiO$_{2}$/LaO interfacial layers the two-dimensional conducting state arises mainly localized within the interfacial MnO layer. The density of states at the Fermi-level can be tuned by ferroelectric polarization reversal. The conducting state coexists with magnetic one mainly arose from Mn atoms.
 %
\end{abstract}

\maketitle


\section{Introduction}

\label{intro}

The discovery of two-dimensional electron gas in 2004 by Ohtomo and Hwang~\cite{ohtomo} boosted a new area of condensed matter physics when it became possible to combine incompatible properties in one material, for instance, superconductivity and magnetism at the LaAlO$_{3}$/SrTiO$_{3}$ interface.~\cite{reyren,bert,luli}
Most previous works on 2DEGs at oxide interfaces were based mainly on SrTiO$_{3}$, in which the interfacial magnetism is very weak since of the nonmagnetic nature of their parent materials. It was also shown that magnetism in such systems is induced by defects. For instance, maximal magnetic moment induced by oxygen vacancy at the interface equals 0.2\,$\mu_{B}$ per Ti atoms neighbouring to the defect.~\cite{piyanzina2019}

To enhance the interfacial magnetism, it is essential to use magnetic insulators as a component of heterostructures in order to support spin-polarized 2DEG. Even more promising issue here is to manipulate it using an electric field, i.e., to realize a converse magnetoelectric (ME) effect. Recently, in Ref.\,\onlinecite{weng} the spin-dependent switching effect was demonstrated on a system based on  perovskites YTiO$_{3}$/PbTiO$_{3}$, where it was suggested to use antiferromagnet (AFM), as a source of magnetism, and ferroelectric (FE) - to manipulate the interfacial states. The use of  A-type AFM material was supported by the fact, that both the FE field effect and the A-type AFM order are layer dependent, and the A-type AFM order will be better coupled with the field effect.~\cite{weng}

In the present work, the LaMnO$_{3}$/BaTiO$_{3}$ (LMO/BTO) heterostructure along the [001] direction will be studied as a model system. 
This system was also investigated previously by Ciucivara \textit{et al.}~\cite{ciucivara2008}, where (LaMnO$ {_3} $)$_{4.5}$/(BaTiO$ { _3} $)$_{4.5}$ superlattice was considered. They reported that starting with A-AFM LaMnO$ {_3} $ the optimization of the superlattice converged to ferromagnetic order. It was concluded that interfaces increase the magnetization and may favour ferromagnetic ordering~\cite{ciucivara2008}. 


Based on density functional theory (DFT) calculations, here we investigate the formed spin-polarized 2DEGs, which is upon the FE switching can change both conducting state and magnetization. 
The essential issue of the presented paper is to understand the impact of polarization inside the BTO slab on the interface electronic and magnetic states. 
In order to do so we investigate  electronic and magnetic properties of BTO/LMO heterostructure without imposed polarisation ($ P_{0} $) and  with ferroelectric polarizations towards the interface ($ P_{down} $) and the surface ($ P_{up} $). 


\section{Computational Method}

\label{method}

The {\it ab initio} calculations were based on density functional  theory (DFT).~\cite{hohenberg1964,kohn1965} Exchange and correlation effects were accounted for by the generalized gradient approximation (GGA) as parametrized by Perdew, Burke, and Ernzerhof (PBE)~\cite{perdew1996}. The Kohn-Sham equations were solved with projector-augmented-wave (PAW) potentials and wave functions~\cite{bloechl1994paw} as implemented in the Vienna Ab-Initio Simulation Package (VASP)~\cite{kresse1996a,kresse1996b,kresse1999}, which is part of the MedeA\textsuperscript{\textregistered} software of Materials Design~\cite{medea}. Specifically, we used a plane-wave cutoff of 400\,eV. The force tolerance was 0.05\,eV/\AA\ and the energy tolerance for the self-consistency loop was $ 10^{-5} $\,eV. The Brillouin zones were sampled using Monkhorst-Pack grids~\cite{monkhorst1976} including $ 5 \times 5 \times 1 $ $ {\bf k} $-points.

A set of calculations was carried out with a \textit{+U} correction applied to Mn 3\textit{d}, Ti 3\textit{d} and La 4\textit{f} states, which can improve the description of the electronic properties of LaMnO$_{3}$ and BaTiO$_{3}$ in bulk and complex configurations, giving the correct splitting of the Mn 3$\textit{d}$ states, as well as the unshift of La 4\textit{f} states~\cite{piyanzina2017}. 
A simplified Dudarev approach was used~\cite{dudarev1998}: the \textit{U}  values of 4\,eV  for Mn,  2\,eV for Ti 3\textit{d}  and 8\,eV for La 4\textit{f} states were applied.

The heterostructures were modelled by a central region of $ {\rm LaMnO_3} $ comprising $ 2 \frac{1}{2} $ unit cells with $ {\rm LaO} $ termination on both sides and varying number of $ {\rm BaTiO_3} $  overlayers with $ {\rm TiO_{2}} $ termination towards the central slab and $ {\rm BaO} $ surface termination also on both sides (the heterointerfaces with other terminations were not converged (same result was found in Ref.~\onlinecite{ciucivara2008}).
The slab model used in the present research guarantees a non-polar structure without any artificial dipoles. Finally, in order to avoid interaction of the surfaces and slabs with their periodic images, $\approx$20~\AA-wide vacuum region was added in accordance with previous works~\cite{cossu2013,piyanzina2017}. The in-plane lattice parameters $ a =  5.709 $\,\AA\ and $b =5.675$\,\AA\ were fixed to the computed values of the bulk $ {\rm LaMnO_3} $ and kept for all subsequent calculations reflecting the stability of the substrate. Whereas all atomic positions of all atoms were fully relaxed.

\section{Results and discussion}
\label{results}

\subsection{LaMnO$_{3}$ and BaTiO$_{3}$ in the bulk and thin film configurations} 
\label{bulk}
First, the parent materials of the heterointerface have been checked separately in the bulk and thin-film geometry in order to ensure the reproducibility of the results obtained by the method and computational parameters used in the present research.

\subsubsection{LaMnO$_{3}$}
Starting from the experimental structure, the lattice constants and atomic positions were fully relaxed.
To obtain the magnetic ground state of LMO, the total energies of A-type, C-type, and G-type antiferromagnetic (AFM) and ferromagnetic (FM) states were calculated. 
\begin{table}[h]
\centering
\caption{Calculated energy (\textit{E}) per unit cell, magnetic moment ($\textit{m}$) per Mn atom and direct band gap ($\varepsilon$) for the bulk LaMnO$_{3}$. Experimental Mn magnetic moment is 3.42-3.87~\cite{hauback,elemans,moussa}, optical band gap equalt to 1.1-1.9\,eV~\cite{chung,tokura,saitoh,yamaguchi} for the bulk LaMnO$_{3}$.}
\label{table1}
\begin{tabular*}{0.28\textwidth}{c | c | c | c  } 
\hline 
Order & \textit{E}, eV & $\textit{m}$, $\mu_{B}$/Mn & $\varepsilon$,\,eV  \\ 
\hline 
FM & -154.93 & 3.778 & 0.404   \\ 
A-AFM & -154.89 & 3.832 & 1.349   \\ 
C-AFM & -154.84 & 3.603 & 1.528   \\ 
G-AFM & -154.85 & 3.542 & 1.652  \\  
\hline 
\end{tabular*}
\end{table}
Our calculation confirms that the FM order has the lowest energy. 
The calculated band gap and local magnetic moment are listed in Table~\ref{table1}, the values for AFM LMO are in good agreement with experimental data also presented in the table caption.

\begin{figure}[h]
 \includegraphics [angle=0,width=9cm]   {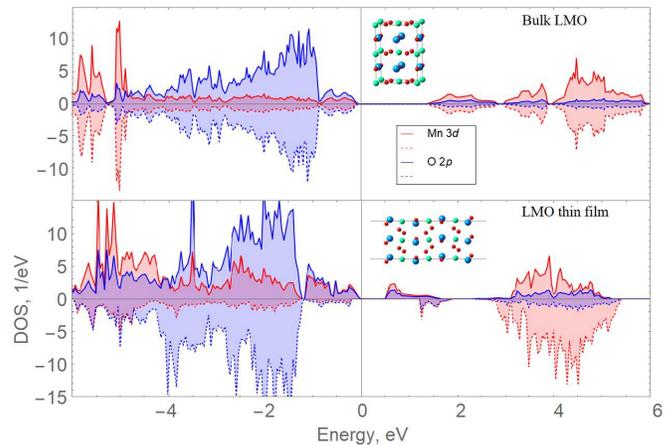}
  \caption{Density of states spectra for the A-AFM bulk and thin-film  LaMnO$_{3}$ along with the corresponding unit cell structures.}
\centering
\label{img:Figure1}
\end{figure}

To realize the spin-dependent switching effect, which was mentioned in the introduction, it was suggested to replace FM material by AFM.~\cite{weng}  The spin-polarized density of states (DOS) plot for the A-AFM LMO in the bulk configuration in presented in Fig.~\ref{img:Figure1}. The bulk LMO is a semiconductor, with an O 2\textit{p} dominated valence band, with Mn 3\textit{d} contribution. The calculated band gap and magnetic moment of the Mn atom equal 1.349\,eV and 3.832\,$\mu_{B}$, respectively. Those values agree well with experimental values of 1.7\,eV and $3.7\pm0.1$\,$\mu_{B}$.~\cite{saitoh} Based on this comparison we concluded that the chosen value of the \textit{U} parameter can yield relatively correctly both the energy gap and the magnetization. Besides, the calculated cell parameters, shown in Table~\ref{table2}, were found to be close to experimental values~\cite{sawada} and also to the previous \textit{ab initio} research, for example Ref.~\onlinecite{ciucivara2008}. 
\begin{table}[h]
\centering
\caption{Calculated lattice constants in \AA\ of bulk LaMnO$_{3}$, bulk  BaTiO$_{3}$ and a rotated by 45$^{\circ}$ along \textit{z}-axis BaTiO$_{3}$ unit cell in order to merge with  the LaMnO$_{3}$ unit cell. The last row lists the BaTiO$_{3}$/LaMnO$_{3}$ supercell lattice constants. Experimental data are presented as well for comparison.}
\label{table2}
\begin{tabular*}{0.3\textwidth}{c | c c c  } 
\hline 
 & a & b & c  \\ 
\hline 
LMO & ~5.709~ & ~5.675~ & ~8.018~   \\ 
Expt.~\cite{sawada} & 5.742 & 5.532 & 7.669 \\ 
\hline 
BTO & 3.986 & 3.986 & 4.014   \\ 
Expt.~\cite{crystal} & 3.992 & 3.992 & 4.036  \\ 
BTO ($\times \sqrt{2}$) & 5.637 & 5.637 & 4.014 \\  
\hline 
Supercell & 5.709 & 5.675 & 50  \\
\hline 
\end{tabular*}
\end{table}

In Fig.~\ref{img:Figure1} the DOS spectrum for the thin film geometry is presented as well in order to check the geometry impact onto the conducting and magnetic properties. The slab was considered to have seven alternating sublayers of MnO$_{2}$ and LaO. In the DOS we see that the band gap decreases due to the Mn 3\textit{d} and O 2\textit{p} states down-shift, but the LMO film remains insulative, whereas the A-AFM order becomes FM in total due to the odd number of MnO layers in the slab.

\subsubsection{BaTiO$_{3}$}

To be ferroelectric, a material must possess a spontaneous dipole moment that can be switched in an applied electric field, i.e., spontaneous switchable polarization. BaTiO$_{3}$ is one of the most well known ferroelectric, which has a ferroelectric polarization in a tetragonal system that forms when cooled from the high-temperature cubic phase, through the Curie temperature, of  T$_{C}$=120$^{\circ}$C. BTO has moderate polarization of 26\,$\mu$C/cm$^{2}$.~\cite{subarao}  Barium titanate has two other phase transitions on cooling further, each of which enhances the dipole moment. The phase which is reached after cooling to 0$^{\circ}$C from tetragonal is orthorhombic. And then there is a rhombohedral below -90$^{\circ}$C. All of these ferroelectric phases have a spontaneous polarisation based to a significant extent on the movement of the Ti atom in the O6 octahedra in different directions. 
Polarization may be defined as the total dipole moment per unit volume.

Calculated energies per unit cells, band gaps, oxygen displacements and polarizations for the cubic, tetragonal and orthorhombic phases are presented in Table~\ref{table3}.
\begin{table}[h]
\centering
\caption{Calculated energy ($\textit{E}$) per unit cell,  c/a ratio, the band gaps ($\varepsilon$), displacement of Ti atoms with respect to the O planes ($\bigtriangleup$) and polarization $\textbf{P}$  of the cubic, tetragonal and orthorombic phases of bulk  BaTiO$_{3}$. Experimental values are given for the tetragonal phase.}
\label{table3}
\begin{tabular*}{0.5\textwidth}{c | c | c | c | c | c } 
\hline 
Phase & $\textit{E}$,\,eV/u.c. & c/a & $\varepsilon$,\,eV & $\bigtriangleup$,\,\AA & $\textbf{P}$,\,$\mu$C/cm$^{2}$  \\ 
\hline
Cubic & -37.33 & 1 & 2.169 & 0 & 0 \\ 
\hline
Tetragonal & -37.29 & 1.007  & 2.249  & 0.13 & 31   \\ 
Expt. & & 1.010\,\cite{shirane} & 3.27\,\cite{wemple}& 0.15\,\cite{shirane} & 26\,\cite{subarao} \\
\hline
Orthorhombic & -37.28 & 1.428 & 2.259 & 0.09 &   \\  
\hline 
\end{tabular*}
\end{table}
All calculated values agree well with experimental data. The experimental band gap is higher than the computed one, but the difference is reasonable for the DFT. As expected, the cubic phase has the lowest energy, but for the purpose of work we are interested in the phases with spontaneous polarization, and in the current research we will focus on the tetragonal BTO structure. The calculated lattice parameters along with experimental ones are listed in Table~\ref{table2}.

\begin{figure}[h]
 \includegraphics [angle=0,width=9cm]   {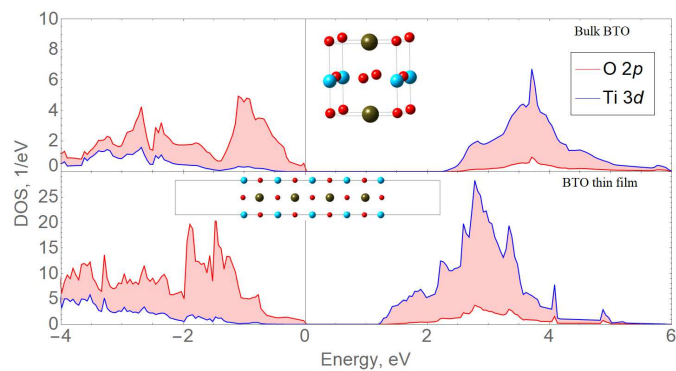}
  \caption{Density of states spectra for the bulk and thin film BaTiO$_{3}$ along with the corresponding unit cell structures.}
\centering
\label{img:Figure2}
\end{figure}

The calculated density of state spectra for the bulk and thin-film BTO are given in Fig.~\ref{img:Figure2}. The thin-film geometry does not change the conducting properties significantly: the overall band gap decreases from 2.249\,eV to 1.221\,eV. At the same time, the spontaneous polarization vanishes  due to the final number of layers.

\subsection{LaMnO$_{3}$/BaTiO$_{3}$ heterostructure}
\label{hetero}
\begin{figure}[h]
 \includegraphics [angle=0,width=8cm]   {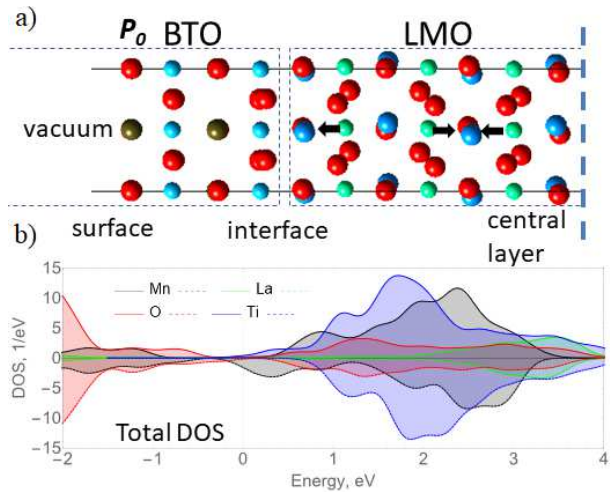}
  \caption{a) The fully optimized half-left structure of LMO/BTO heterointerface. The supercell is displayed without   the vacuum region, which was ~20\AA. Arrows indicate magnetic moments directions in the LMO slab. b) Corresponding atom-resolved density of states.}
\centering
\label{img:cell}
\end{figure}

In order to merge  BTO with LMO so that the polarization is parallel to the easy axis of antiferromagnet, the BTO unit cell has to be rotated by  45$^{\circ}$ along \textit{z}-axis. As listed in Table\,\ref{table2}  $a_{BTO}\times\sqrt{2}$ are very close to the $a_{LMO}$ and $b_{LMO}$ cell parameters. The resulted supercell is presented in Fig.~\ref{img:cell}\,a, where the left half of the unit cell is presented without full vacuum region. The structure was fully optimized.

\begin{figure}[h]
 \includegraphics [angle=0,width=8cm]   {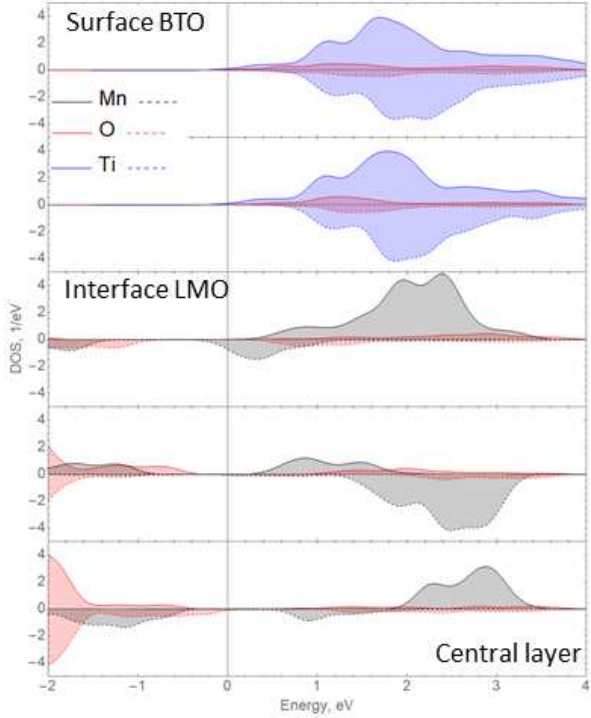}
  \caption{The layer-resolved density of states spectra for the LMO/BTO heterostructure without imposed polarization. The optimized structure of LMO/BTO heterointerface without imposed polarization. Full LMO/BTO unit supercell is displayed but without the vacuum region, which was ~20\,\AA. Arrows indicate magnetic moments directions in the LMO slab. }
\centering
\label{img:Figure3}
\end{figure}

Performed structural optimization resulted in the insignificant shift of Ti atoms out of oxygen planes. Within the interfacial layer $\Delta$\,z$_{Ti-O}$ distances equal to -0.046\,\AA\ and 0.247\,\AA , within the surface layer shifts are less significant 0.049\,\AA\ and 0.101\,\AA , respectively.  Except for the one, all the movements of Ti are towards the surface, which leads to the total polarization predominantly towards the surface. Such a structural reconstruction leads to the electronic rearrangement as reflected in the DOS spectrum (Fig.~\ref{img:cell}\,b and Fig.~\ref{img:Figure3}). 

The  layer-resolved DOS (shown in Fig.\ref{img:Figure3}) reveals, that  the Fermi level is mainly crossed by states filled by electrons of the interfacial Mn atoms of LMO, implying the interfacial 2DEG. DOS at the Fermi level for the second and third layers inside the LMO slab is about zero.
The mechanism behind arising conductivity might well be the following: due to the structural displacements of oxygens towards the interface, the ferroelectric polarization towards the surface of the slab appears in the BTO film. And the total maximum of electric field strength is directed from LMO towards the BTO surface.  As a consequence, all electrons will be attracted by the lowest interfacial BTO layer and confined near the interface.

Besides, the layer-resolved DOS clearly demonstrates the magnetic moments' distribution within LMO. As was mentioned in the beginning, the LMO was chosen to be in A-AFM phase, but due to the odd number of MnO layers in total LMO is FM. During the optimization, it remains A-AFM distribution of magnetic moments as shown in Fig.~\ref{img:Figure3}. 

\begin{figure}[h]
 \includegraphics [angle=0,width=8cm]   {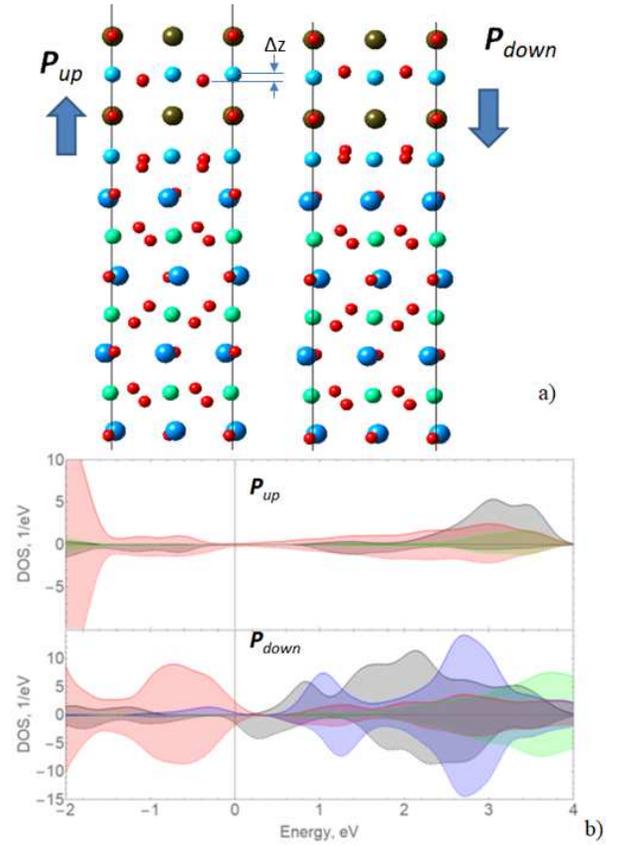}
  \caption{a) Half structures of LMO/BTO supercells with modelled imposed polarization i.e., dipoles directed toward the surface (P$_{up}$) and towards the interface (P$_{down}$) as shown by blue arrows. b) The corresponding density of states spectra.}
\centering
\label{img:Figure4}
\end{figure}

In order to realize the polarization to be in positive or negative directions with respect to $\textit{z}$ axis, the Ti atoms were moved away from the oxygen planes up or down by $\approx$0.33 and BTO slabs were frozen during the optimization, whereas the central atoms of LMO were all fully relaxed.
The two considered supercells are shown in Fig.~\ref{img:Figure4} with corresponding DOSes. 

%
%

It is clearly observed from  Fig.~\ref{img:Figure4} that there is an impact of polarization onto the electronic states. Even though the system remains conductor for both up and down cases, the main contribution at the Fermi level changes from Mn spin down in the \textbf{P}$_{0}$ (Fig.~\ref{img:cell})  to the oxygen contribution for \textbf{P}$_{up}$ case and oxygen and increased Mn contribution for  \textbf{P}$_{down}$ case. Basically, the middle case for \textbf{P}$_{up}$ is the same direction of dipoles but with increased intensity of $\Delta$\,z$_{Ti-O}$ shifts. This increased shift if Ti atoms out of oxygen planes gave rise to the upshift of Mn spin-down component and consequently the decrease of DOS at the Fermi level.
In contrast, the lower \textbf{P}$_{down}$ case corresponds to the opposite shift of Ti atoms within the BTO slab, so that dipoles are directed towards the interfaces. That changes the DOS spectrum drastically: there is an overall shift of states towards the Fermi level and a much higher intensity of DOS. 

At the same time, we have checked, that the distribution of magnetic moments  within the MnO layers does not change significantly. However, the total magnetization is the lowest for P$_{down}$ and the highest for P$_{0}$. So, the magnetoelectric coupling takes place, but as not as pure as in Ref.~\onlinecite{weng} for YTiO$_{3}$/PbTiO$_{3}$, where the interfacial magnetic moments turn to zero for P$_{up}$. One possible reason, except for the different compounds used, is different supercells: in the mentioned research there is even number of MnO layers and, consequently, the A-AFM ordering in the YTiO$_{3}$ slab. Besides, the slab is not symmetric and no vacuum was added in contract to our supercell. 
%
%

\section{Summary}
\label{summary}
 
In summary, by means of \textit{ab initio} calculations, we have demonstrated  that a spin-polarized 2DEG occurs in the LMO/BTO system without imposed polarization, localized mainly in the MnO  layers with a maximum at the interface. Therefore, the coexistence of magnetism in a 2DEG, i.e., a spin-polarized 2DEG, is presented in the LMO/BTO heterointerface. Arising conducting state occurs due to the structural deformations primarily  within the interfacial TiO$_{2}$ layer, leading to the electronic reconstructions and downshift of Mn states in the conduction band.
Then, we have shown that the combination of FE polarization and antiferromagnetism can affect the spin-polarized 2DEG, in particular, imposed polarization may change the conducting state.

\begin{acknowledgments}
The reported study was funded by the Russian Scientific Foundation according to research project No. 21-12-00179. Computing resources were provided by the Laboratory of Computer design of new materials at Kazan Federal University.
\end{acknowledgments}

\end{document}